\documentclass[3p,twocolumn]{elsarticle}
\usepackage{graphicx}
\usepackage{amsmath}
\usepackage{amssymb}
\usepackage{color}
\usepackage{soul}
\usepackage[usenames,dvipsnames]{xcolor}
\usepackage{verbatim}
\usepackage{url}
\usepackage{bm}
\usepackage{multirow}
\usepackage{booktabs}
\usepackage{rotating}
\usepackage{lipsum}
\usepackage{gensymb}
\usepackage{lineno,hyperref}
\modulolinenumbers[5]

\begin{document}

\begin{frontmatter}
	
	\title{Identification of the nature of excitons in PPDT2FBT using electroabsorption spectroscopy}
	
	\author [1] {Subhamoy Sahoo}
	\author [2] {Rajdeep Dhar}
	\author [2]{Sanjoy Jena}
	\author [2] {Soumya Dutta}
	\author [2]{Debdutta Ray}
	\author [1]{Jayeeta Bhattacharyya}
	\address[1]{Department of Physics, Indian Institute of Technology Madras, Chennai 600036, India.}
	\address[2]{Department of Electrical Engineering, Indian Institute of Technology Madras, Chennai 600036, India.}
	\cortext[mycorrespondingauthor]{Corresponding author}
	\ead{jayeeta@iitm.ac.in}
	
	
	\begin{abstract}
		
		Electroabsorption (EA) measurements can be used to identify the type of excitons contributing to the absorption spectra of semiconductors which have applications in optoelectronics. However, the inferences from the EA measurement greatly depend on the method of fitting and extraction of parameters from the measured spectra. We deconstruct the absorption spectrum by fitting multiple Gaussians and obtain the relative contribution of first and second derivative of each absorption band in EA spectrum, which gives indication of the Frenkel, charge transfer or mixed nature of the excitons involved. We check the applicability of the method for pentacene which is widely used and well studied organic semiconductor. We report EA measurements of poly[(2,5-bis(2-hexyldecyloxy)-phenylene)-alt-(5,6-difluoro-4,7-di(thiophen-2-yl)benzo[c]-[1,2,5]-thiadiazole)] (PPDT2FBT). Our analysis shows that besides the feature around 3.07 eV, which is strongly Frenkel-like, most of the absorption bands for PPDT2FBT are mixed states, having relatively high charge transfer contributions. Since charge transfer excitons have higher dissociation efficiencies, we infer PPDT2FBT to be a promising candidate for photovoltaic applications. 
		
	\end{abstract}
	
	
\end{frontmatter}


\section{Introduction}
The optical and electrical properties of the organic semiconductors are primarily excitonic, due to their high binding energy of about few hundreds of meV \cite{pope1999}. When excited, organic semiconductors exhibit two types of excitons, namely, Frenkel and charge transfer (CT) \cite{Hoffmann2000} . For a Frenkel exciton, the hole and electron are localized on the same molecule, whereas the CT excitons extend over nearest or next to nearest molecules \cite{pope1999,Wang2019}. There has been a lot of interest in understanding the CT excitons, as they play a vital role in determining the performance of organic solar cells due to their more efficient charge separation \cite{qin2019}. On the other hand, in organic light emitting diodes (OLED), where radiative recombination of the excitons is required, the involvement of Frenkel excitons is important \cite{varene2012}. Recent studies have shown that CT excitons can also contribute to improving the efficiency of OLED by enhancing the singlet excitons population by reverse intersystem crossing \cite{zhang2013, wu2015}. Typically, the absorption spectra of organic semiconductors show multiple features associated with different excitonic transitions in the material. Thus, it is important to understand the type of excitons giving rise to the individual features to optimize the efficiencies of optoelectronic devices. 

We present our study on the nature of exciton of poly[(2,5-bis(2-hexyldecyloxy)-phenylene)-alt-(5,6-difluoro-4,7-di(thiophen-2-yl)benzo[c]-[1,2,5]-thiadiazole)] (PPDT2FBT) which has emerged as a potential candidate for organic solar cell. PPDT2FBT is a low band gap (1.76 eV) organic semiconductor polymer having highest occupied molecular orbital (HOMO) and lowest unoccupied molecular orbital (LUMO) at -5.45 eV and -3.69 eV, respectively \cite{nguyen2014, ahn2019simple} . Though, a relatively newer material, PPDT2FBT and its fullerene derivatives (PC$_{70}$BM) have shown potential applications in solar energy harvesting due to their wide absorption spectrum, long term thermal stability and high hole mobility nearly independent of thickness up to $\sim$ 1 $\mu$m \cite{nguyen2014, ko2016photocurrent, koh2017enhanced}. The thick structure helps to attenuate the incident sunlight without damaging the fill factor \cite{ko2016photocurrent, shin2019ultra, dayneko2019indoor}, providing a pathway to defect free film processing over large areas \cite{zhao2017}. Large area solar cell, using PPDT2FBT as an active material, has shown encouraging power conversion efficiencies \cite{ko2016photocurrent}. Though a lot of work have been done on PPDT2FBT from the device perspective, the optical properties are yet to be explored. In this report, we distinguish the Frenkel and CT like excitonic transitions in PPDT2FBT which is an important study for optimal use of the polymer for opto-electronic applications.

Different methods like DFT calculations \cite{petelenz1999}, binding energy estimation \cite{park2018,qi2013} have been reported to determine the excitonic properties of organic semiconductors. \textit{L. Sebastian et al.}\cite{sebastian1981} used electroabsorption (EA) spectroscopy to study the CT transitions in tetracene and pentacene and later this approach has been employed by other groups as well \cite{petelenz1988, bernardo2014}.

In EA spectroscopy, the change in absorption of a material due to an externally applied electric field is measured. Being a modulation measurement, the EA spectrum typically resembles derivative of the zero-field absorption spectrum $[A(E)]$.  In case of Frenkel excitons, the EA spectrum replicates the first derivative of the absorption spectrum $[A^{'} (E)=(dA(E))/dE]$, arising due to the change in polarizability upon excitation. For CT transitions, however, the EA spectrum is more of second derivative $[A^{''} (E)=(d^2 A(E))/(dE^2)]$ in nature, originating from the significant change in dipole moment of the spatially separated CT exciton \cite{sebastian1981, jalviste2007}.  For overlapping peaks in the absorption spectrum, integral method analysis of EA spectra has been used by \textit{Kattoor et al.}\cite{kattoor2018integral}. In our work, we addressed each transition appearing in the absorption spectrum of PPDT2FBT individually by using Gaussian fits to the transitions \cite{Legaspi2018}. The EA spectrum was then fitted as a whole, with weighted $A^{'} (E)$ and $A^{''} (E)$ contributions from the individual transitions \cite{Legaspi2018}. This allowed to distinguish the excitonic nature of the different overlapping peaks in the absorption spectra of PPDT2FBT.

We present EA measurement and analysis for pentacene thin films to establish the viability of our analysis.  We chose pentacene since it is a well studied system  which has been extensively used in organic thin film transistor  and has regenerated interest recently, due to the observation of singlet fission in the material \cite{Wilson2013}. Then, EA measurements and analysis were done for PPDT2FBT which gave insight on the types of excitons contributing to its absorption spectrum required for improving its performance in solar cell applications.

\section{Experimental Section}
\begin{figure}[!h]
 	\centering
 	\includegraphics[width = 0.75\linewidth]{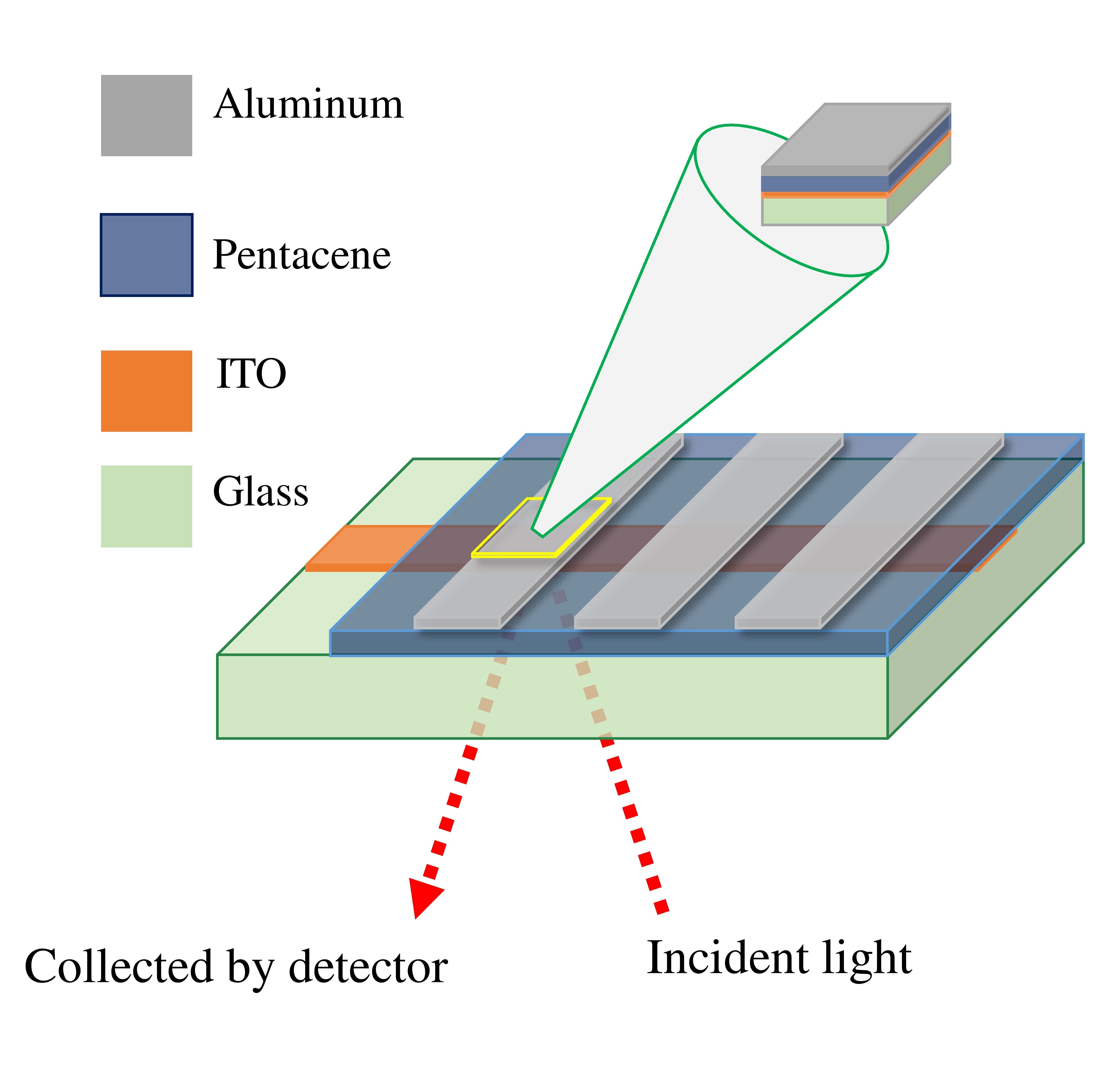}
 	\caption{The schematic of ITO/Pentacene/Al device. PPDT2FBT has also similar device structure. The dotted line shows the path of incident and reflected light. The zoomed section shows the cross section of the device }
 	\label{Fig:Device_structure}
 \end{figure}

 A 100 nm thick pentacene film was deposited on a cleaned ITO coated glass substrate, by thermal evaporation at $4.7 \times 10^{-7}$ mbar pressure at a rate of 2 $\AA$/s. Then, aluminum (2 mm $\times$ 10 mm) fingers of about 80 nm thickness were deposited on top of it (Figure \ref{Fig:Device_structure}), forming the cathode of the device. For the PPDT2FBT device, PPDT2FBT solution having concentration of 40 mg/ml, was spin coated onto a patterned ITO (anode), at 1000 rpm, and annealed at $130^{0}$C for 10 min. The average film thickness of PPDT2FBT was estimated to be 135 nm by a Brooker profilometer. The EA measurements were done in the reflection geometry. In all the devices, ITO acted as the front electrode, through which the light was made incident on the organic semiconductor at an angle of 45 deg and reflected by the bottom Al film.
 
 The EA spectrum was measured using Xe lamp as light source, which was coupled to a monochromator. The light was incident on the semiconductor film, though the ITO/glass substrate, reflected back from the Al electrode and  measured using a Si photodiode, by phase sensitive detection using a lock-in amplifier. A function generator was used to bias the sample with both DC and sinusoidal AC (at 1.3 kHz) voltage. The TTL output of the function generator was fed as a reference to the lock-in amplifier. The DC background ($R$) of the samples were obtained from measurements where the biasing voltage was turned off and the incident light was chopped by a mechanical chopper to which the lock-in amplifier was locked. As the signal was measured by lock in amplifier, a factor $\frac{\sqrt{2} \pi}{2}$  is required to convert the rms value to amplitude. The change in reflectance ($\Delta R$) was measured at the first harmonic of the modulation frequency so a factor $\sqrt{2}$ is required to convert rms value to the amplitude. EA signal at first harmonic is proportional to the applied AC and DC fields (supplementary section \textbf{S1}). The applied AC and DC fields for pentacene device were $1.06 \times 10^{5}$ V/cm and $4.60 \times 10^{5}$ V/cm, respectively. For PPDT2FBT device these values were $2.61 \times 10^{4}$ V/cm and $2.67 \times 10^{5} $ V/cm respectively. Thin films of each sample, grown on the glass substrate at the same growth condition as the devices, were used to measure the respective absorption spectra in transmission mode.
 
\section{Results and Discussion}
The change in the absorbance $[\Delta A(E)]$ of a semiconductor due to an externally applied electric field could be expressed as a linear combination of the zero field absorption $[A(E)]$ and its first $[A^{'} (E)]$ and second $[A^{''} (E)]$ derivatives w.r.t. the photon energy $(E)$ \cite{jalviste2007}, as stated by equation \ref{equn:1}.
\begin{equation}
	\Delta A(E)=\left[rA(E)+p \ \frac{dA(E)}{dE} + q\ \frac{d^2 A(E)}{dE^2}\right]
	\label{equn:1}
\end{equation}

The coefficients of the derivative terms ($r$, $p$ and $q$) provided the information about the type of exciton. The coefficient of the zeroth order derivative ($r$) was significant for anisotropic distribution of ensemble of transition dipole moments in the material \cite{jalviste2007}. It depended on the value of the ground state polarizability along the direction of the transition dipole moment. The first derivative contribution, defined by $p$, was a combined effect of the orientation of the molecular dipole along the applied field and the linear Stark effect due to the change in dipole moment associated with the absorption. The field induced broadening of the absorption spectrum due to quadratic Stark shift gave rise to the second derivative term, of strength $q$ \cite{jalviste2007}. The values of $p$ and $q$, derived from fitting the EA spectrum, would thus reveal the dependence of that transition on the change in polarizability $(\Delta P)$ and dipole moment $(\Delta \mu)$, indicating the nature of the associated exciton. The coefficients $p$ and $q$ obtained from first harmonic EA spectrum, were related to the change in polarizability $(\Delta P)$ and dipole moment $(\Delta \mu)$ given by equation \ref{equn:2} and \ref{equn:3} (supplementary section \textbf{S1}).

\begin{equation}
	p = \frac{1}{2} \ \Delta P |2 F_{AC} F_{DC}^{'}|
	\label{equn:2}
\end{equation}

\begin{equation}
q = \frac{1}{6} \ \Delta \mu^{2} |2 F_{AC} F_{DC}^{'}|
\label{equn:3}
\end{equation}

Here $F_{AC}$ was the applied AC field and $F_{DC}^{'}$ comprised of the applied DC bias ($F_{DC}$) and the built-in field ($F_{BI}$) generated within the film. This method was applicable for both small organic molecules and conjugated polymers, with the assumption that the inter molecular, inter-chain and intra-chain interactions were ignored \cite{jalviste2007}. For conjugated polymers, we considered that the EA signal originated from the effective change in polarizability and dipole moment of the molecules.

The absorption spectrum typically consisted of the superposition of Gaussians corresponding to the different transitions as expressed in equation \ref{equn:4} \cite{Legaspi2018}. The EA spectrum was fitted with a linear combination of zeroth, first and second derivatives of each absorption band weighted by factors given by $r_i$, $p_i$ and $q_i$ (equation \ref{equn:5}).

\begin{equation}	
	A(E) = \sum_{i} a_{i} \exp[-((E-E_i)/w_i )^2 ] = \sum_{i} A_{i} (E)
	\label{equn:4}
\end{equation}

\begin{equation}	
\Delta A(E) = [r_{i}A_{i} (E)  + p_{i} A_{i}^{'}  + q_{i} A_{i}^{''}]
\label{equn:5}
\end{equation}

Here $A_i (E)$ was the $i^{th}$ Gaussian band of the absorption spectrum, having amplitude, peak position and width given by $a_i$, $E_i$, and $w_i$ respectively. $A_i (E)$, $A_i^{'} (E)$ and $A_i^{''} (E)$ were the zeroth, first and second derivatives of the $i^{th}$ Gaussian band with respect to energy ($E$) and $r_i$, $p_i$ and $q_i$ were the coefficients, respectively.
When measured in the reflection geometry, the $\Delta A$ of an organic layer of thickness $d$ could be expressed in term of relative change in reflectance $(\Delta R/R)$ by equation \ref{equn:6}, assuming that there was negligible transmission loss and $\Delta R$ in presence of electric field is much smaller than the zero field reflectance $(R)$ (supplementary section \textbf{S2}).

\begin{equation}	
\Delta A(E) = -0.13 \times \frac{\Delta R}{R}
\label{equn:6}
\end{equation}

The absorption spectrum was deconstructed by Gaussian functions (equation \ref{equn:4}) and parameters $a_i$, $E_i$ and $w_i$ were extracted. The fitting was done using commercial software, where the data was fitted iteratively using predefined Levenberg-Marquardt algorithm. From the analytical functional form of derivatives of a Gaussian, the first and second derivatives of each absorption bands were calculated. The EA spectrum was then fitted with equation \ref{equn:7} to obtain $r_i$, $p_i$ and $q_i$ values. The details of the fitting is discussed in the supplementary information (section \textbf{S3}).

\begin{equation}	
-\frac{\Delta R}{R} =\frac{1}{0.13} \times  [r_{i}A_{i} (E)  + p_{i} A_{i}^{'}  + q_{i} A_{i}^{''}]
\label{equn:7}
\end{equation}

The value of the parameters obtained from the fitting and FWHM of individual bands, are shown in the Table \ref{tab:table_pentacene_constarin_fit_0.13_corrected_2_corrected}-\ref{tab:table_ppdt2fbt_2_0.13_corrcted_2_corrected}. The FWHM typically told about the energy spread of the transitions, which was quite large for amorphous materials and depended on the origin of the transition. We found that the FWHM was larger for PPDT2FBT compared to pentacene. In our case the peak positions and FWHM of the individual features were important as they indicated the extent of overlap between the nearest bands.

\subsection{Analysis of EA spectrum of Pentacene}

 \begin{figure}
	\centering
	\includegraphics[width = \linewidth]{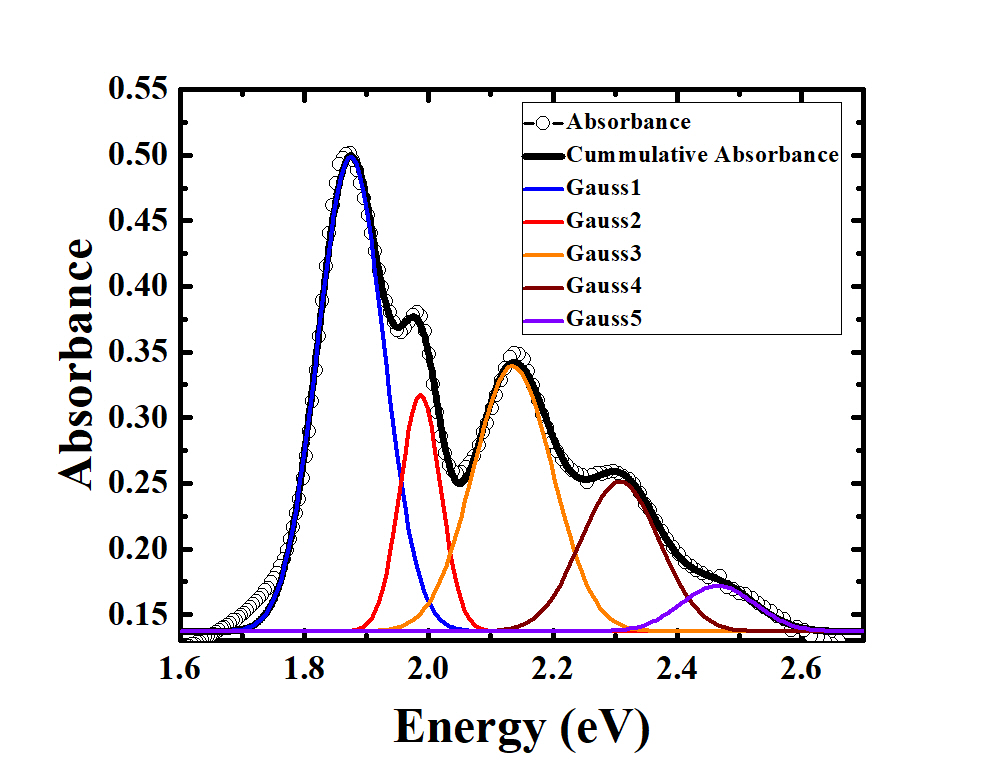}
	\caption{Absorption spectrum of Pentacene. Open circles and solid black line represent the measured spectrum and the fitted spectrum respectively. The coloured Gaussian bands represent different absorption bands. }
	\label{Fig:Pentacene_Abs}
\end{figure}

\begin{figure}
	\centering
	\includegraphics[width = 0.95\linewidth]{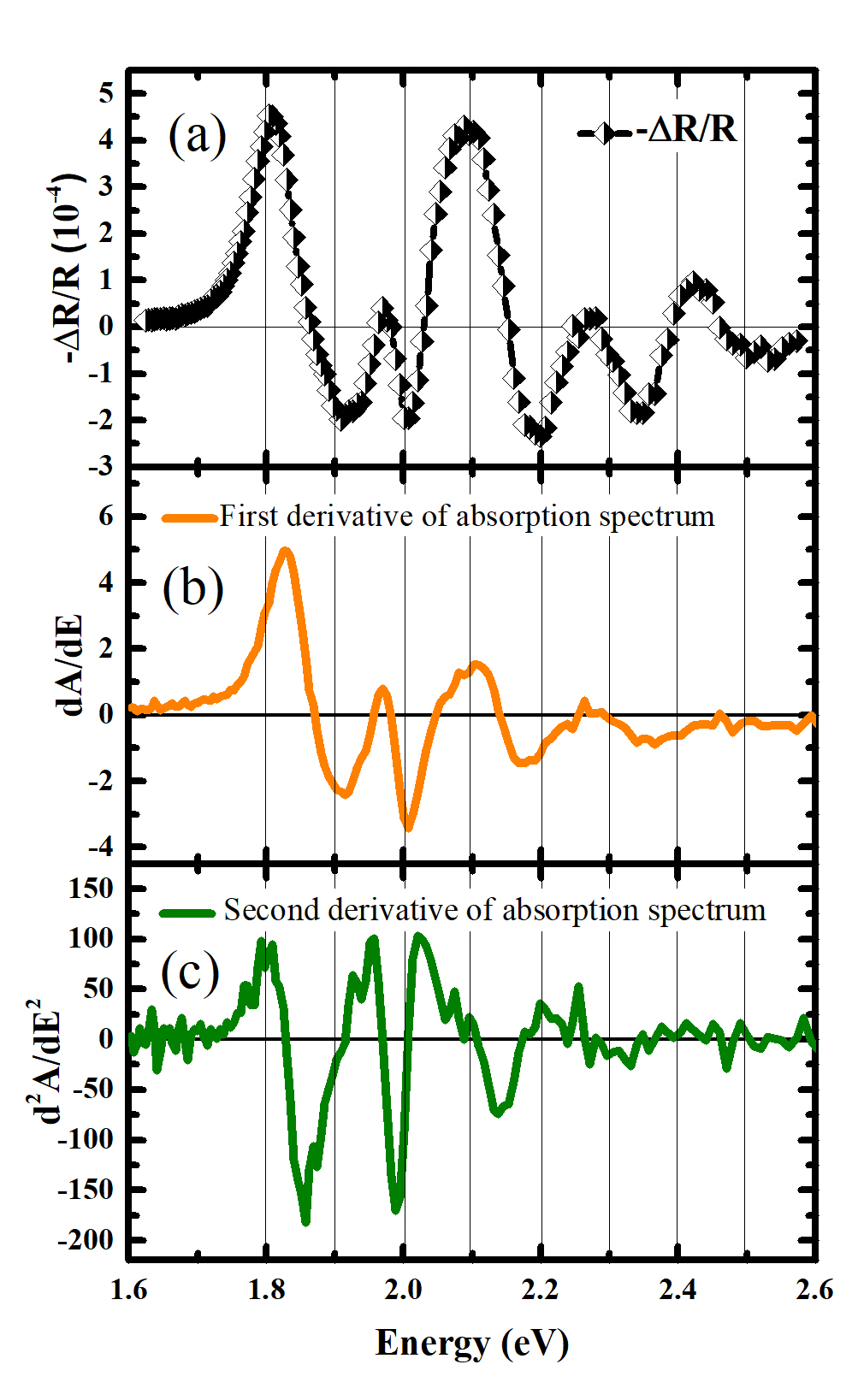}
	\caption{(a) Electroabsorption spectrum of pentacene, (b) first and (c) second derivative of absorption spectrum, obtained numerically}
	\label{Fig:Pentacene_EA_comparison}
\end{figure}

\begin{table*}
	\caption{\label{tab:table_pentacene_constarin_fit_0.13_corrected_2_corrected}The position, FWHM of the absorption bands and the coefficients of the derivatives extracted from the fitting of Pentacene and corresponding $\Delta P$ and $\Delta \mu$   }
	\centering
	\begin{tabular}{cccccccc}
		\hline
		Peak &	FWHM &	$r$  &	$p$  &	$q$  &	$|q/p|$ & $\Delta P$ & $\Delta \mu$\\ 
		position (eV) &	(meV) &	  &	(eV) &	(eV$^2$) &	(eV) & ($\AA^3$) & (D)\\ 
		\hline
		$1.87$ &	$125.4$ &$ 0.00 $ & $1.19\times 10^{-5}$ &$8.21\times 10^{-8}$ &	$6.9 \times 10^{-3}$ & $35.2$ & $1.08$ \\
		$1.99$ &	$76.8 $ &	$0.00$	 & $1.06\times 10^{-5}$ &$5.33\times 10^{-8}$ &	$5.0 \times 10^{-3}$ & $31.6$ & $0.86$\\
		$2.14$  &	$150.9$ &	$0.00$	 &$ 3.02\times 10^{-5}$ &$ 0.00 $	 &  $0.00$ & $89.2$ & $0$ \\
		$2.30 $ &	$148.8 $&	$1.51\times 10^{-4}$  &$2.54\times 10^{-5}$ &$8.74\times 10^{-7}$ &	$3.44 \times 10^{-2}$ & $74.5$ & $3.52$\\
		$2.47$  &	$141.8$ &	$0.00$	 & $2.94\times 10^{-5}$ &$0.00$ &	$0.00$ & $86.8$ & $0$\\
		\hline
		
	\end{tabular}
\end{table*}

\begin{figure}[h!]
	\centering
	\includegraphics[width = \linewidth]{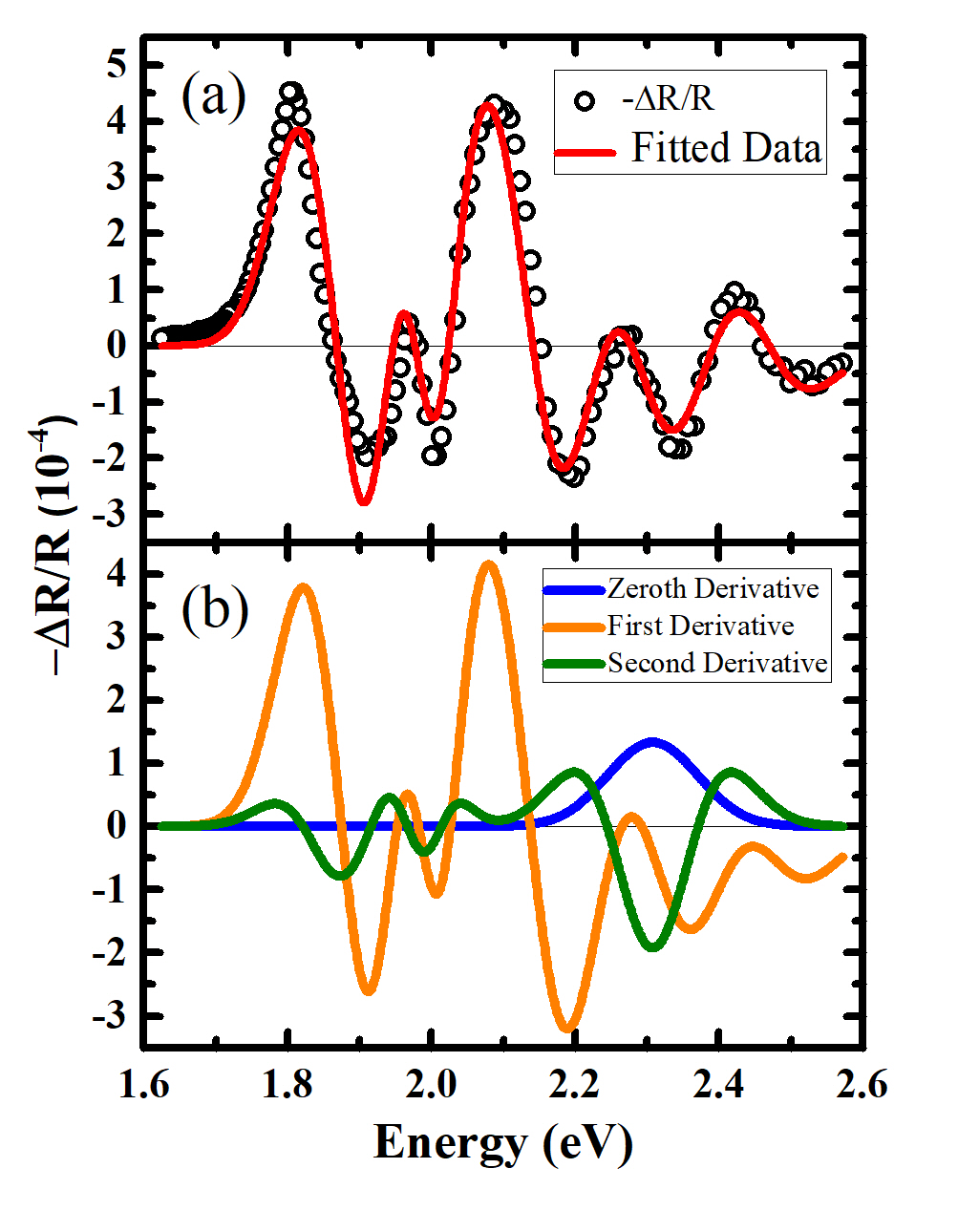}
	\caption{(a)Measured EA spectrum of Pentacene (open circle) and the data fitted by the linear combination of the derivatives (solid line) and (b) the contribution of different derivatives to the EA spectrum.}
	\label{Fig:Pentacene_EA_fit}
\end{figure}

The absorption spectrum of pentacene thin film (Figure \ref{Fig:Pentacene_Abs}) consisted of five distinct peaks and was reconstructed by five symmetric gaussians. 
The comparison of the measured EA spectrum for pentacene with the first and second derivatives of the entire absorption spectrum is shown in Figure \ref{Fig:Pentacene_EA_comparison}. Though, there seemed to be a good likeliness of the EA spectrum with the first derivative curve, most of the peak positions did not match, as were indicated by the dotted lines. We therefore fitted the EA spectrum with equation \ref{equn:7} (Figure \ref{Fig:Pentacene_EA_fit}a). The fitted values gave the q/p ratio and the change in polarizability and dipole moment for the individual transitions are listed in Table - \ref{tab:table_pentacene_constarin_fit_0.13_corrected_2_corrected}. The contribution of different derivatives in EA spectrum is shown in the Figure \ref{Fig:Pentacene_EA_fit}b.

The bands at 2.14 eV and 2.47 eV had Frenkel characteristic as the second derivative components for these bands were zero. The change in polarizability for these bands were found to be 89.2 and 86.8 $\AA^3$, respectively.  As seen from Table – \ref{tab:table_pentacene_constarin_fit_0.13_corrected_2_corrected}, the q/p ratio for peak at 1.87 eV and 1.99 eV were $6.9 \times 10^{-3}$ and $5.0 \times 10^{-3}$ which implied its strong Frenkel-like nature. This was also indicated by the good match of the EA signal around 1.9 eV with the first derivative spectrum in Figure \ref{Fig:Pentacene_EA_comparison}. The corresponding change in polarizabiliy due to optical transition associated with these bands were 35.2 and 31.6 $\AA^3$, respectively. The much higher value of the q/p ratio for peak at 2.30 eV indicated that it had relatively higher CT characteristics \cite{sebastian1981}. The change in dipole moment for this band was 3.52 D compared to the bands at 1.87 eV and 1.99 eV for which $\Delta \mu$ were 1.08 and 0.86 D respectively.

The dominant Frenkel characteristic of the absorption bands at 1.99 eV and 1.87 eV was in agreement with the previously reported results\cite{sebastian1981, haas2010field}. The other two absorption band at higher energy (2.14 eV and 2.30 eV) was assigned to be charge transfer states. But, from our analysis, we found that the state at 2.30 eV was a mixed-state, having comparatively higher CT characteristic and the state at 2.14 eV had Frenkel characteristic. This might arise from the segment wise fitting of the EA spectrum, done by the \textit{L. Sebastian et al.}\cite{sebastian1981} where overlapping of the bands were neglected. The EA spectrum was reproduced by only first derivative of absorption spectrum by \textit{Haas et al.}\cite{haas2010field} which might be partially correct since the linear combination of the derivatives gave a better fit.

\subsection{Analysis of EA of PPDT2FBT}

\begin{figure}
	\centering
	\includegraphics[width = \linewidth]{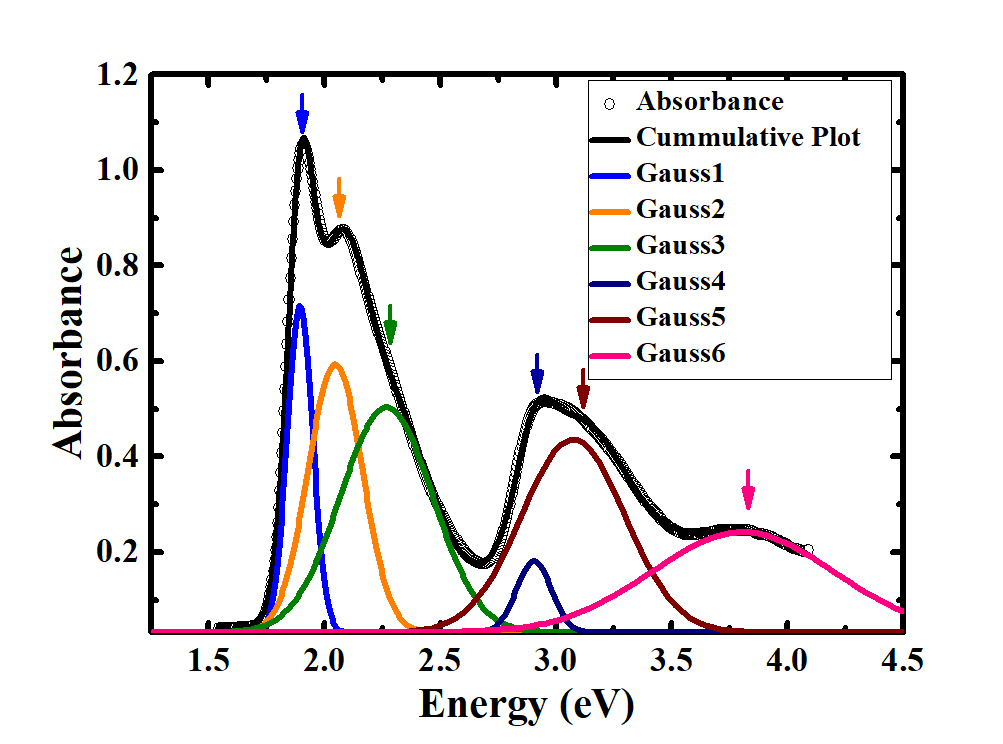}
	\caption{Absorption spectrum of PPDT2FBT. Open circles and solid black line represent the measured spectrum and the fitted spectrum respectively. The coloured Gaussian bands represent different absorption bands. }
	\label{Fig:PPDT2FBT_Abs}
\end{figure}

\begin{table*}[h!]
	\caption{\label{tab:table_ppdt2fbt_2_0.13_corrcted_2_corrected}The position, FWHM of the absorption bands and the coefficients of the derivatives extracted from the fitting of PPDT2FBT and corresponding $\Delta P$ and $\Delta \mu$ }
	\centering
	\begin{tabular}{cccccccc}
		\hline
		Peak &	FWHM &	$r$  &	$p$  &	$q$  &	$|q/p|$ & $\Delta P$ & $\Delta \mu$\\ 
		position (eV) &	(meV) &	  &	(eV) &	(eV$^2$) &	(eV) & ($\AA^3$) & (D)\\ 
		\hline
		$1.89$&	$129.4$&	$0.00$ &	$-1.40\times 10^{-7}$&	$1.85\times 10^{-8}$&	$1.32 \times 10^{-1}$ & $-2.9$ & $1.35$\\
		$2.05$&	$275.7$&	$0.00$ &	$5.34\times 10^{-7}$&	$1.60\times 10^{-7}$&	$2.99 \times 10^{-1}$ & $11.0$ & $3.98$\\
		$2.27$&	$469.7$&	$6.92 \times 10^{-6}$ &	$1.37\times 10^{-6}$&	$ 1.35 \times 10^{-7} $&	$9.85 \times 10^{-2}$ & $28.2$ & $3.66$\\
		$2.90$&	$177.6$&	$0.00$&	$-5.10\times 10^{-8}$&	$ 5.37\times 10^{-8}$&	$1.05$ & $-1.1$ & $2.30$\\
		$3.07$&   $525.3$&  $0.00$ &  $1.14\times 10^{-6}$&   $ 0.00 $  &  	$0.00$ & $23.4$ & $0$\\
		\hline
		
	\end{tabular}
\end{table*}

\begin{figure}[h!]
	\centering
	\includegraphics[width = \linewidth]{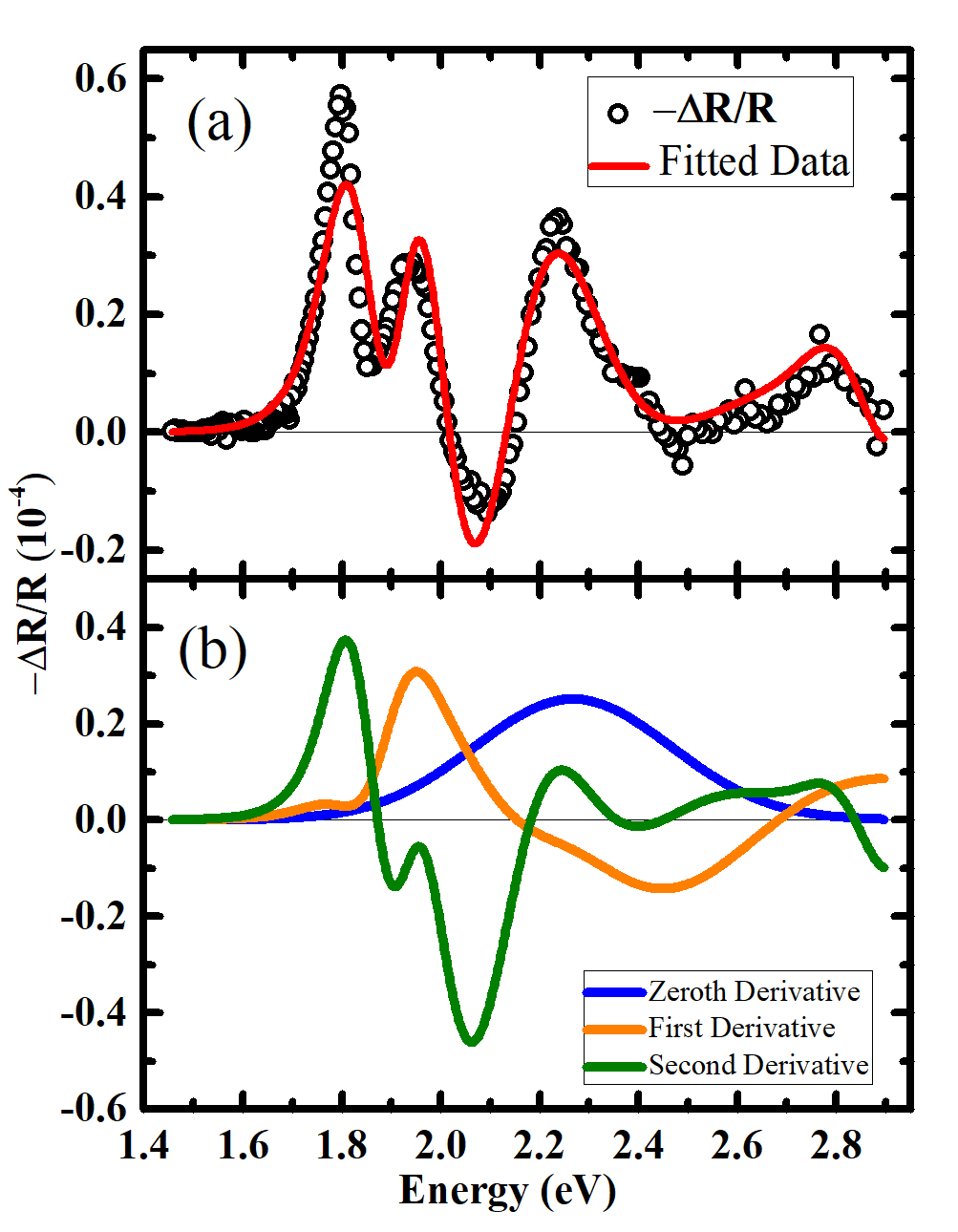}
	\caption{(a) Measured EA spectrum of PPDT2FBT (open circle) and the data fitted by the linear combination of the derivatives (solid line). (b) The contribution of zeroth, first and second order derivatives to the EA spectrum}
	\label{Fig:PPDT2FBT_EA_fit}
\end{figure}

The absorption spectrum of PPDT2FBT ranges over almost entire visible wavelength (Figure \ref{Fig:PPDT2FBT_Abs}), and consists of multiple features. Being one of the relatively newer materials, the nature of the excitons contributing to the absorption spectrum in PPDT2FBT is not well known. We, therefore, used this methodology to investigate the excitonic properties using EA spectroscopy. The absorption spectrum was fitted with six gaussians (Table-\ref{tab:table_ppdt2fbt_2_0.13_corrcted_2_corrected}) which was found to be the minimum number of peaks (shown by arrows in Figure \ref{Fig:PPDT2FBT_Abs})required to reconstruct the absorption spectrum. The measured EA spectrum of PPDT2FBT is shown in the Figure \ref{Fig:PPDT2FBT_EA_fit}a. When fitted with equation \ref{equn:7}, the obtained values of the coefficients were listed in the Table – \ref{tab:table_ppdt2fbt_2_0.13_corrcted_2_corrected} and the contribution of different derivatives in the EA spectrum was shown in the Figure \ref{Fig:PPDT2FBT_EA_fit}b. In the fitting, the band at 3.81 eV was not considered as this had negligible contribution in the spectral range of measured EA spectrum, to minimize the number of free parameters and avoid overfitting.

For the absorption band at 3.07 eV the q/p ratio suggested that the associated excitons were Frenkel in nature. The change in polarizability associated with this transition was 23.4 $\AA^3$. The transitions at 1.89 eV, 2.05 eV,  2.27 eV and 2.90 eV showed much higher value of the ratio indicating mixed nature of these transitions, with more CT component. The maximum change in dipole moment was observed for the band at 2.05 eV (3.98 D). The change in dipole moment for other two bands at 1.89 and 2.90 eV were 1.35 D and 2.30 D respectively. The change in polarizabiliy was smaller for these mixed states compare to the band at 2.27 eV which had a high $\Delta \mu$ (3.66 D) along with comparatively high polarizability of 28.2 $\AA^3$

\section{Conclusion}
We presented a line shape based analysis of electroabsorption spectra to identify the nature of excitons in organic semiconductors. For pentacene, the features around 1.87, 1.99, 2.14 and 2.47 eV originated from a strong Frenkel-like exciton whereas the absorption band at 2.30 eV had higher CT characteristic . We reported EA measurement and analysis for PPDT2FBT. By fitting the EA spectrum, we identified a strong Frenkel-like transition around 3.07 eV, while other contributing transitions were more CT-like. The CT nature being dominant for most of the optically active transitions in PPDT2FBT, implied higher probability of exciton dissociation. This makes PPDT2FBT a potential candidate for photovoltaic applications.



\bibliography{ref}

\end{document}


%
	
	
	

\section*{\Large{Supplementary Information}}

\vspace*{0.25 cm}

\section*{\Large{Identification of the nature of excitons in PPDT2FBT using electroabsorption spectroscopy}}
Subhamoy Sahoo, Rajdeep Dhar, Sanjoy Jena, Soumya Dutta, Debdutta Ray, Jayeeta Bhattacharyya

\vspace*{1 cm}

\section*{S1: The EA signal at the first harmonic of modulation frequency}	

The change in absorption coefficient varies quadratically with the applied field as shown by equation \ref{equn:1} \cite{sebastian1981}.

\begin{equation}
\Delta \alpha (E) = [A_\chi \alpha (E) + B_\chi \alpha^{'} (E) + C_\chi \alpha^{''} (E)] \times F^2 
\label{equn:1}
\end{equation}	

The absorption coefficient and the absorbance is related by $\alpha d = 2.303 A$ where $d$ is the thickness of the organic layer. So, equation \ref{equn:1} can be expressed in terms of absorbance.

\begin{equation}
\Delta A (E) = [A_\chi A (E) + B_\chi A^{'} (E) + C_\chi A^{''} (E)] \times F^2 
\label{equn:2}
\end{equation}	

In modulation measurement an AC bias, superimposed on a DC background is to the sample. If the angular frequency of the sinusoidal modulation is $\omega$  then the total field will be  $[F_{AC}  \sin (\omega t) +F_{DC}^{'} ]$. So,

\begin{equation}
\Delta A (E) = [A_\chi A (E) + B_\chi A^{'} (E) + C_\chi A^{''} (E)] \times [F_{AC}  \sin (\omega t) +F_{DC}^{'} ]^2
\label{equn:3}
\end{equation}	

So, the signal at first harmonic of the modulation frequency will be 

\begin{equation}
\Delta A (E) = [A_\chi A (E) + B_\chi A^{'} (E) + C_\chi A^{''} (E)] \times [2 F_{AC}  F_{DC}^{'} \sin (\omega t) ]
\label{equn:4}
\end{equation}

The DC field ($F_{DC}^{'}$) is the resultant of the applied field ($F_{DC}$) and built in field ($F_{BI}$). We measured the EA spectrum in reverse bias so  $F_{DC}^{'}=F_{DC}+F_{BI}$.

\begin{equation}
\Delta A (E) = [A_\chi A (E) + B_\chi A^{'} (E) + C_\chi A^{''} (E)] \times [2 F_{AC}  \ (F_{DC} + F_{BI}) \sin (\omega t) ]
\label{equn:5}
\end{equation}

Equation \ref{equn:5} can be written as the following which is equation 1 in the manuscript.
\begin{equation}
\Delta A (E) = [r A (E) + p A^{'} (E) + q A^{''} (E)] 
\label{equn:6}
\end{equation}

So, the change in polarizability ($\Delta P$) and dipole moment ($\Delta \mu$) associated with a transition are related to the coefficients $p$ and $q$ by following equations \cite{sebastian1981,iimori2018influence,iimori2019fluorescence,bernardo2014}.

\begin{equation}
p = \frac{1}{2} \ \Delta P |2 F_{AC} F_{DC}^{'}|
\label{equn:6.1}
\end{equation}

\begin{equation}
q = \frac{1}{6} \ \Delta \mu^{2} |2 F_{AC} F_{DC}^{'}|
\label{equn:6.2}
\end{equation}

\section*{S2: Relation between $\Delta A$ and $\frac{\Delta R}{R}$, measured by lock in amplifier	}

\begin{figure}[h!]
	\centering
	\includegraphics[width=0.5\textheight]{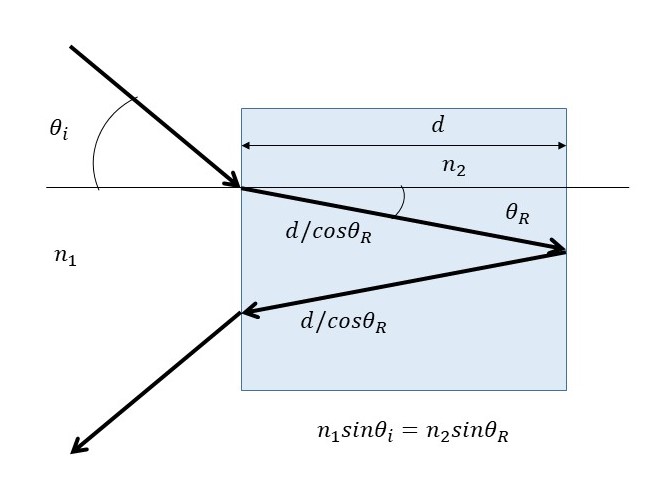}
	\caption{Effective thickness of the organic layer in reflection geometry}
	\label{Fig:1}
\end{figure}

In reflection mode, the effective thickness of the device will be $2d/\cos(\theta_R)$ where $d$ is the thickness and $\theta_R$ is the angle of refraction, can be found from Snell's law i.e. $n_1 \sin\theta_i=n_2 \sin\theta_R$ (Figure \ref{Fig:1}). In this case $\Delta \alpha $ and the change in reflectance in presence of field $\left( \frac{\Delta R}{R}\right)_A $ are expressed by equation \ref{equn:7}.
\begin{equation}
\Delta \alpha  = -\frac{\cos\theta_R}{2d} \left( \frac{\Delta R}{R}\right)_A
\label{equn:7}
\end{equation}

The absorbance (A) is measured in transmission mode, so $\alpha d = 2.303 A$. So,

\begin{equation}
\Delta A  = -\frac{\cos\theta_R}{2} \frac{1}{2.303} \left( \frac{\Delta R}{R}\right)_A 
\label{equn:8}
\end{equation}	

In experiment, $R$ is measured by lock in amplifier by chopping the incident light by a mechanical chopper. So, R will have shape close to the square wave. And $\Delta R$ is measured by applying a sinusoidal bias to the sample. The lock in amplifier measures the rms value of the signal so in this case $\Delta R_{rms} =\frac{1}{\sqrt{2}} \Delta R_A$ and $R_{rms} =\frac{1}{\sqrt{2}} \frac{2}{\pi} R_A$

\begin{equation}
\frac{\Delta R}{R} = \frac{\Delta R_{rms}}{R_{rms} } = \frac{\pi}{2} \left( \frac{\Delta R}{R}\right)_A
\label{equn:9}
\end{equation}

We can rewrite the equation \ref{equn:8} by rearranging the terms.

\begin{equation}
\Delta A = - \frac{\cos \theta_R}{2} \times \frac{1}{2.303} \times \frac{2}{\pi} \times  \frac{\Delta R}{R}
\label{equn:10}
\end{equation}	
\textbf{$\cos \theta_R/2$}: Term represents effective thickness\\
\textbf{$1/2.303$}: For converting $\Delta \alpha$ to $\Delta A$\\
\textbf{$2/\pi$}: For converting rms value to amplitude\\

For pentacene and PPDT2FBT the value of the refractive index is in the range of 1.54-2.0 and 1.45 - 2.95 respectively in the wavelength range of our interest. For $45^{0}$ angle of incidence, $\cos \theta_R$ approximately can be taken as a constant, having values 0.92. So, equation \ref{equn:10} can be
written as 	

\begin{equation}
\Delta A \approx - 0.13 \times  \frac{\Delta R}{R} 
\label{equn:11}
\end{equation}

\section*{S3: The procedure of fitting the EA spectrum}	

The EA spectrum is first compares with the derivatives of the zero field absorption spectrum. Then it is checked that whether the linear combination of different derivatives with equal weightage can fit the spectrum. As this method of fitting can not generate a good fit, the linear combination of derivatives with different weightage is used for it. We settle down for the set of parameters which has minimum number of free parameters and have reasonably good fit. During fitting we constrain the coefficient of the second derivative ($q$) to be positive. Different combination of parameters is tried for fitting and the best fitting with minimum number of free parameters are presented in the manuscript. To fit the spectrum, an analytical form the derivatives are used which are obtained from the reconstruction of the absorption spectrum.


\bibliography{ref}